\documentclass[pra,twocolumn,showpacs,longbibliography]{revtex4-1}

\usepackage[latin1]{inputenc}
\usepackage{exscale}
\usepackage{graphicx}
\usepackage{amsmath}
\usepackage{amssymb}
\usepackage{color}
\usepackage{bbold}
\usepackage{easybmat}
\usepackage{rotating}

\newcommand{\ket}[1]{\vert #1\rangle}
\newcommand{\bra}[1]{\langle #1\vert}

\newcommand{\ee}{\mathrm{e}}
\newcommand{\ii}{\mathrm{i}}
\newcommand{\dd}{\mathrm{d}}

\newcommand{\an}[1]{\hat{#1}}
\newcommand{\cre}[1]{\hat{#1}^\dag}

\begin{document}

\title{Exploiting boundary states of imperfect spin chains for high-fidelity state transfer}

\author{M. Bruderer}
\author{K. Franke}
\author{S. Ragg}
\author{W. Belzig}
\affiliation{Fachbereich Physik, Universität Konstanz, D-78457 Konstanz, Germany}

\author{D. Obreschkow}
\affiliation{International Centre for Radio Astronomy Research, The University of Western Australia, 35 Stirling Hwy, Crawley WA 6009, Australia}

\date{\today}

\begin{abstract}
We study transfer of a quantum state through $XX$ spin chains with static
imperfections. We combine the two standard approaches for state transfer
based on (i) modulated couplings between neighboring spins throughout the
spin chain and (ii) weak coupling of the outermost spins to an unmodulated
spin chain. The combined approach allows us to design spin chains with
modulated couplings and localized boundary states, permitting high-fidelity
state transfer in the presence of random static imperfections of the
couplings. The modulated couplings are explicitly obtained from an exact
algorithm using the close relation between tridiagonal matrices and
orthogonal polynomials [Linear Algebr. Appl. \textbf{21}, 245 (1978)]. The
implemented algorithm and a graphical user interface for constructing spin
chains with boundary states (\textsc{spinGUIn}) are provided as Supplemental
Material.
\end{abstract}



\maketitle


\section{Introduction}\vspace{-5pt}

Spin chains have attracted much attention in recent years (for reviews
see~\cite{Bose-CP-2007,Kay-IJQI-2010}) because of their ability either to
act as quantum communication
channels~\cite{Bose-PRL-2003,Christandl-PRL-2004,Wojcik-PRA-2005} or to
generate highly entangled states for quantum
computation~\cite{Yung-PRA-2005,Clark-NJP-2005,Clark-NJP-2007}. The use of
spin chains for both of these tasks has been considered in the context of
various physical systems: Implementations of spin chains may connect
nitrogen-vacancy registers in diamond~\cite{Yao-PRL-2011} or entangle
internal states of an array of ultracold atoms confined to an optical
lattice~\cite{Clark-NJP-2005}. Arrays of capacitively coupled flux qubits
have also been shown to be suited for quantum state
transfer~\cite{Lyakhov-NJP-2005}.

In particular, spin chains of $XX$~type~\cite{Lieb-AP-1961} can be used as a
quantum channel, i.e., a quantum state (qubit) placed at one end of the
chain can be perfectly transferred to the other end as a result of the
coherent time evolution. Two different strategies have been suggested to
achieve perfect state transfer (PST) in this context. The first approach
relies on modulated couplings between neighboring spins throughout the spin
chain~\cite{Christandl-PRL-2004,Yung-PRA-2005}. More precisely, the quality
of the state transfer depends on the energy spectrum of the spin chain and
suitable couplings for PST are obtained by solving an inverse eigenvalue
problem (IEP). The prototypical couplings between spins for PST correspond
to a linear spectrum~\cite{Christandl-PRL-2004}, also considered relevant to
the dynamics of electrons in $N$-level
systems~\cite{Eberly-PRA-1977,Bialynicka-PRA-1977,Cook-PRA-1979,Shore-1979}.

The second approach is based on weak couplings of the outermost spins to the
rest of the otherwise unmodulated spin
chain~\cite{Wojcik-PRA-2005,Wojcik-PRA-2007,Zwick-PRA-2012}. In this
weak-coupling limit the dynamics of the spin chain reduces to an effective
two- or three-level system consisting of states localized at the boundaries
of the spin chain (boundary states) and state transfer is a consequence of
Rabi-type oscillations~\cite{Shore-1979,Wojcik-PRA-2007}. PST is only
achieved in the limit of vanishing couplings of the outermost spins of the
chain. The advantage of this approach is that transfer does not depend on
details of the spin chain, i.e., no modulation of inner spin couplings is
required.

The primary aim of this paper is to combine these two strategies. We
construct spin chains with modulated couplings; however, these are chosen
such that state transfer takes place mainly through boundary states. The
resulting spin chains therefore have two qualities: they permit PST through
a perfectly engineered chain, and their dynamics, involving only a few
states, is robust to small variations of the spin couplings. To construct
these spin chains we start from the energy spectrum and add nearly zero
eigenvalues, which under specific conditions entails boundary states.

The couplings of the spin chains are explicitly obtained by solving the IEP
with a numerically stable algorithm developed by de~Boor and
Golub~\cite{deBoor-LAA-1978}. We describe the algorithm adapted to the
specific problem of PST through spin chains and provide its implementation
as \textsc{matlab} code. In addition, we supply a graphical user interface
for designing spin chains with boundary states, called \textsc{spinGUIn}, as
Supplemental Material~\cite{supplemental}.

As a proof of principle, we compare transfer fidelities of different spin
chains with static imperfections of the spin couplings, similar to the
analysis in Refs.~\cite{DeChiara-PRA-2005,Zwick-PRA-2011,Zwick-PRA-2012}.
The imperfections are modeled as relative fluctuations of the couplings
drawn from a uniform distribution. The resulting distributions of the
transfer fidelities reveal that chains with boundary states are more
resilient to imperfections. This is reflected in more instances of
high-fidelity transfer through the spin chain.

The structure of this article is as follows: In Sec.~II we recall the basics
of state transfer through spin chains. In Secs.~III and IV we give necessary
details of the two strategies for achieving state transfer using modulated
couplings and weakly coupled end spins. Furthermore, we describe the
algorithm by de~Boor and Golub, which may be applied to the modulated
coupling strategy. In Sec.~V we present the combined approach and apply it
to specific examples, including an analysis of their transfer fidelity in
the presence of imperfect couplings. We end with the conclusions in Sec.~VI.


\section{State transfer}

To start with, we recall the basics of state transfer through a
one-dimensional $XX$ spin chain with the Hamiltonian
\begin{equation}\nonumber
    \hat{H}_S = \frac{1}{2}\sum_{j=1}^{N-1} b_j\left(\an{\sigma}^x_{j}\an{\sigma}^x_{j+1}
    + \an{\sigma}^y_{j+1}\an{\sigma}^y_{j}\right) + \frac{1}{2}\sum_{j=1}^N a_j \left(\mathbb{1}+\an{\sigma}^z_{j}\right)\,.
\end{equation}
Here, $b_j$ are the spatially dependent spin couplings between neighboring
sites, $a_j$ are local external fields, and $\hat{\sigma}^{x},
\hat{\sigma}^{y}, \hat{\sigma}^{z}$ are the Pauli matrices. It is convenient
to map the spin Hamiltonian $\hat{H}_S$ to a one-dimensional fermionic
hopping model using the Jordan-Wigner transformation~\cite{Lieb-AP-1961},
which yields the equivalent Hamiltonian
\begin{equation}\nonumber
    \hat{H}_F = \sum_{j=1}^{N-1} b_j\left(\cre{c}_{j}\an{c}_{j+1} + \cre{c}_{j+1}\an{c}_{j}\right) + \sum_{j=1}^N a_j \cre{c}_{j}\an{c}_{j}\,.
\end{equation}
The operators $\cre{c}_{j}$ ($\an{c}_{j}$) create (annihilate) a fermion at
site~$j$ and obey the usual anticommutation relations. Since the Hamiltonian
$\hat{H}_F$ commutes with the total number operator
$\hat{n}=\sum_{j=1}^N\cre{c}_{j}\an{c}_{j}$, the Hilbert space can be
decomposed into subspaces $\mathcal{H}_n$ corresponding to different total
fermion numbers $n$. For transferring a single qubit we restrict our
considerations to the subspace $\mathcal{H}_0 \oplus \mathcal{H}_1$. The
subspaces $\mathcal{H}_0$ and $\mathcal{H}_1$ are spanned by the vacuum
state $\ket{\mathrm{vac}}$ and the single-fermion Fock states $\ket{j} =
\cre{c}_{j}\ket{\mathrm{vac}}$, respectively.

For state transfer the spin chain is initialized in the vacuum state. The
qubit is written into the first spin of the chain at time $t=0$, so that the
state of the spin chain is $\ket{\psi(0)} = c_0\ket{\mathrm{vac}} +
c_1\ket{1}$. Subsequently, the qubit is transferred under the coherent
evolution $\hat{U}(t) = \ee^{-\ii\hat{H}_Ft}$ to the spin at site $N$ after
the transfer time $t = \tau$, where it can be read out. In the ideal case we
have $\ket{\psi(\tau)} = \hat{U}(\tau)\ket{\psi(0)} = c_0\ket{\mathrm{vac}}
+ c_1\ket{N}$, i.e., PST is achieved. The unitary operator $\hat{U}(\tau)$
and therefore the Hamiltonian $\hat{H}_F$ have to fulfill certain conditions
to ensure PST. Since the vacuum state has a trivial time evolution, the
conditions only apply to the Hamiltonian for the subspace $\mathcal{H}_1$,
which in the single-fermion basis $\{\ket{j}\}$ is given by the tridiagonal
matrix
\begin{equation}\nonumber
    H_N = \left(
  \begin{array}{ccccc}
    a_1 & b_1 & 0 & \cdots & 0\\
    b_1 & a_2 & b_2 & \cdots & 0 \\
    0 & b_2 & a_3 & \cdots & 0 \\
    \vdots & \vdots & \vdots & \ddots & b_{N-1} \\
    0 & 0 & 0 & b_{N-1} & a_N
  \end{array}
\right)\,.
\end{equation}


\section{Modulated couplings}

The approach based on modulated couplings between neighboring spins relies
on two conditions of $H_N$ (for details see~\cite{Yung-PRA-2005}). First,
the matrix $H_N$ has to be symmetric along the antidiagonal, i.e., the
entries of $H_N$ fulfill the condition $a_{N-j+1} = a_{j}$ and $b_{N-j} =
b_{j}$. The matrix $H_N$, being symmetric along the antidiagonal, is said to
be persymmetric. As a result of the reflection symmetry, the eigenvectors
$\ket{\lambda_k}$ of the matrix $H_N$ have definite parities. Moreover, if
the eigenvalues $\lambda_k$ are in increasing order then the eigenvectors
$\ket{\lambda_k}$ change parity alternatively, i.e.,~the mirror-inverted
eigenstates $\ket{{\overline\lambda}_k}$ satisfy the relation
$\ket{{\overline\lambda}_k} = (-1)^k\ket{\lambda_k}$ upon assuming that even
(odd) $k$ label even (odd) eigenstates $\ket{\lambda_k}$.

Second, for PST the eigenvalues $\lambda_k$ have to fulfill the condition
\begin{equation}\label{cond}
    \ee^{-\ii\lambda_{k}\tau} = (-1)^k\ee^{\ii\Phi}
\end{equation}
for a constant transfer time $\tau$ and phase $\Phi$. In fact, if the
initial single-fermion state $\ket{\varphi(0)}\in\mathcal{H}_1$ is expanded
in terms of eigenvectors as $\ket{\varphi(0)} = \sum_{k} c_k
\ket{\lambda_k}$ with constant coefficients $c_k$, then the state at
time~$\tau$ is given by $\ket{\varphi(\tau)} = \sum_{k} c_k\ee^{-\ii
\lambda_k\tau} \ket{\lambda_k}$. On the other hand, since
$\ket{\varphi(\tau)}$ is the mirror-inverted state of $\ket{\varphi(0)}$, by
assumption we have $\ket{\varphi(\tau)} = \ee^{\ii\Phi}\sum_{k} c_k
\ket{\overline{\lambda}_k} = \ee^{\ii\Phi}\sum_{k}
c_k(-1)^{k}\ket{\lambda_k}$. A comparison between the two expressions for
$\ket{\varphi(\tau)}$ then indeed yields the condition in Eq.~(\ref{cond}).
The generalization by global phase factor $\ee^{\ii\Phi}$ can be made since
the phase $\Phi$ can be compensated for, e.g.,~by applying a constant
external field $a_j = -\Phi/\tau$ for all $j$.

Thus, finding the Hamiltonian $H_N$ for PST reduces to an IEP, namely,
calculating the couplings $a_j$ and $b_j$ for a given sequence of
eigenvalues $\lambda_k$ that fulfill the condition in Eq.~(\ref{cond}) for a
fixed $\tau$. A convenient choice is to set the transfer time $\tau$ to the
fixed value $\pi$ so that the eigenenergies $\lambda_k$ take integer values.
Other transfer times are obtained by rescaling the spectrum by an overall
energy scale.

\subsection*{Solving the IEP}

We now describe the algorithm developed by de~Boor and
Golub~\cite{deBoor-LAA-1978} for solving the IEP in the case where $H_N$ is
persymmetric and all eigenvalues $\lambda_k$ are distinct. Solving the
IEP based on continued fractions has been suggested recently in
Ref.~\cite{Wang-PRA-2011} for achieving PST, and similarly in the context of
electric circuit theory~\cite{Sussman-JFI-1982}. We chose the algorithm by
de~Boor and Golub, which was in part motivated by
Ref.~\cite{Hochstadt-LAA-1974}, because of its clarity and straightforward
numerical implementation.

We start with basic definitions related to orthogonal polynomials and
tridiagonal matrices. We denote by $H_j$ the left principal submatrix, which
is formed by deleting the last $N-j$ rows and columns of $H_N$. Further, we
introduce the polynomials $p_j(x) = \det(x-H_j)$, with $j=1,\ldots,N$, and
define $p_0 = 1$ and $p_{-1} = 0$. Clearly $p_j(x)$ are the characteristic
polynomials of the matrices $H_j$, and in particular, $p_N(\lambda_k) = 0$
for the eigenvalues $\lambda_k$. It then follows directly from Laplace's
formula for the expansion of determinants that the polynomials $p_j(x)$
satisfy the three-term recursion relation
\begin{equation}\label{three}
    p_j = (x-a_j)p_{j-1} - b^2_{j-1}p_{j-2}\,.
\end{equation}

Next we introduce the discrete scalar product $\langle\;,\:\rangle$ defined
as
\begin{equation}\label{scalar}
    \langle f,g\rangle = \sum_k w_k\,f(\lambda_k)g(\lambda_k)
\end{equation}
for any polynomials $f$ and $g$ up to degree $N$. As shown in
Ref.~\cite{deBoor-LAA-1978}, the polynomials $p_j$ are orthogonal with
respect to the scalar product in Eq.~(\ref{scalar}), i.e., $\langle
p_i,p_j\rangle = 0$ for $i\neq j $, provided that the spectrum-dependent
weights are defined by $w_k = \left|\dd p_N(x)/\dd x\right|^{-1}$ evaluated
at $x=\lambda_k$. Using the expression $p_N(x) = \prod_k (x-\lambda_k)$ for
the characteristic polynomial, one finds the explicit form for the weights:
\begin{equation}\label{weights}
    w_k = \bigg|\prod_{p\neq k}(\lambda_k-\lambda_p)\bigg|^{-1}\,.
\end{equation}

The orthogonality of the polynomials $p_j$ and the recurrence relation make
it possible to express the coefficients $a_j$ and $b_j$ solely in terms of
$p_j$ and $p_{j-1}$. By taking the scalar product with $p_{j-1}$ on both
sides of Eq.~(\ref{three}) one obtains
\begin{equation}\label{ak}
    a_j = \frac{\langle x p_{j-1},p_{j-1}\rangle}{\langle p_{j-1},p_{j-1}\rangle}\,.
\end{equation}
Similarly, taking the scalar product with $p_{j-2}$ and $p_{j}$ on both
sides of Eq.~(\ref{three}) yields $b_{j-1}^2\langle p_{j-2},p_{j-2}\rangle =
\langle x p_{j-1}, p_{j-2}\rangle$ and $\langle p_j,p_j\rangle = \langle
p_j,x p_{j-1}\rangle$, respectively. Hence using the property of the scalar
product $\langle xf,g\rangle = \langle f,x g\rangle$, one finds
\begin{equation}\label{bk}
    b_j = \frac{||p_{j}||}{||p_{j-1}||}
\end{equation}
with the norm $||f|| = \sqrt{\langle f,f\rangle}$.

The algorithm is based on the key observation that the polynomials $p_j$
and coefficients $a_j$ and $b_j$ can be determined recursively starting with
the polynomials $p_{-1} = 0$ and $p_0 = 1$. The required weights $w_k$,
which specify scalar product in Eq.~(\ref{scalar}), are readily calculated
from the eigenvalues $\lambda_k$. Thus the algorithm for solving the IEP
consists of the following steps:
\begin{enumerate}\setcounter{enumi}{0}
  \item Calculate the weights $w_k$ for the given $\lambda_k$ from
      Eq.~(\ref{weights}). Subsequently repeat steps 2--4~for increasing
      $j$ starting with $j=1$.
  \item Calculate the coefficient $a_j$ from Eq.~(\ref{ak}).
  \item Find the values of $p_j(x)$ at $x = \lambda_k$ from
      Eq.~(\ref{three}).
  \item Calculate the coefficient $b_j$ from Eq.~(\ref{bk}).
\end{enumerate}
For odd $N$ the steps have to be repeated up to $j = (N+1)/2$ and for even
$N$ up to $j=N/2$.

As noted by de~Boor and Golub, the algorithm provides a stable means of
computing the entries of $H_N$. However, the maximum length of the spin
chain $N$ is limited by floating point under- or overflow in the weights or
the values of the polynomials. To ameliorate this problem during execution
of the algorithm it is advisable to scale the spectrum $\{\lambda_k\}$ into
the interval $[-1,1]$ with appropriate rescaling of the resulting
coefficients $a_j$ and $b_j$. This prevents the weights $w_k\sim
1/\Delta\lambda^{N-1}$ ($\Delta\lambda$ being the typical difference between
any two of the $\lambda_k$) from becoming smaller than the typical floating
point precision. Moreover, the polynomial terms $p_j(\lambda_k)$ in the
calculation are potentially large without scaling since they take values of
the order of $\lambda_k^{(N+1)/2}$ for $j=(N+1)/2$. With this proviso
the algorithm yields accurate results for spin chains with lengths up to a
few hundred spins, as found by testing against the exact solutions for the
linear and cosine spectrum (defined in Sec.~V).

Note that the algorithm does not involve approximations and can be used to
obtain exact analytic results for the coefficients $a_j$ and $b_j$. For
instance, a useful analytic result following from Eq.~(\ref{ak}) and based
on symmetry arguments is that all $a_j$ vanish if the eigenvalues
$\lambda_k$ are symmetrically distributed around zero (cf.~Lemma~5
in~\cite{Kay-IJQI-2010}); the converse is also true~\cite{Shore-1979}.


\section{Weakly coupled end spins}

The second approach relies on weak couplings of the outermost spins (end
spins) to the rest of the chain, i.e., the couplings $b_1$ and $b_{N-1}$ are
considerably smaller than the otherwise arbitrary spin couplings $b_j$ with
$j=2,\ldots,N-2$. To analyze this type of spin chain we can therefore use
perturbation theory in the couplings $b_1$, $b_N$. Accordingly, we partition
the Hamiltonian into two parts $H_N = H_0 + V$ with
\begin{equation}\nonumber
V = \left(
\begin{BMAT}(rc){cc}{cc}
\begin{BMAT}(rc){ccc}{ccc}
        0 & b_1 &\\
        b_1 & 0 &\\
        & &
    \end{BMAT} & \mbox{\huge 0} \\
\mbox{\huge 0} & \begin{BMAT}(rc){ccc}{ccc}
 & &\\
& 0 & b_{N-1}\\
& b_{N-1} & 0
\end{BMAT}
\end{BMAT}
\right)
\end{equation}
and $H_0 = H_N - V$, where for simplicity we assumed that the local external
fields $a_j$ vanish.

As customary, we introduce the eigenvalues $\xi_k$ and eigenstates
$\ket{\xi_k}$ of $H_0$, i.e., $H_0\ket{\xi_k} = \xi_k\ket{\xi_k}$. Before
applying perturbation theory we make the following observations: First, the
eigenvalues $\xi_k$ of $H_0$ are symmetrically distributed around zero.
Therefore, if the number of spins $N$ is odd then the Hamiltonian $H_0$ has
an eigenstate $\ket{\xi_0}$ with $\xi_0 = 0$, i.e., a zero mode.  This mode
has the property that all odd components in the position basis $\{\ket{j}\}$
vanish identically and that the even components have alternating signs.
Second, $H_0$ has two additional zero modes regardless of $N$ which are
localized at sites $1$ and $N$, i.e., the two boundary states
$\ket{\xi_1}\equiv\ket{1}$ and $\ket{\xi_N}\equiv\ket{N}$ are zero modes of
$H_0$. In sum, the $\xi_k = 0$ subspace is spanned by the states
$\ket{\xi_0},\ket{\xi_1},\ket{\xi_N}$ for odd $N$ and by the states
$\ket{\xi_1},\ket{\xi_N}$ for even $N$.

We determine the evolution of the state $\ket{\varphi(t)}$ initially in
state $\ket{\xi_1}\equiv\ket{1}$ by using time-dependent perturbation
theory~\cite{Langhoff-RMP-1972}. To this end $\ket{\varphi(t)}$ is expanded
in the basis $\{\ket{\xi_k}\}$ as
\begin{equation}\nonumber
    \ket{\varphi(t)} = \sum_k c_k(t)\,\ee^{-\ii\xi_k t}\ket{\xi_k}
\end{equation}
where $c_k(t)$ are time-dependent coefficients with initial conditions
$c_k(0) = \delta_{1k}$. Inserting $\ket{\varphi(t)}$ into the Schrödinger
equation yields
\begin{equation}\nonumber\label{system}
    \ii\frac{\partial}{\partial t}c_p(t) = \sum_k V_{p k}\,c_k(t)\,\ee^{-\ii(\xi_k - \xi_p)t}
\end{equation}
with the matrix elements $V_{pk} = \bra{\xi_p}V\ket{\xi_k}$. The coupled
equations for the $c_k(t)$ can be solved approximately by separation of time
scales~\cite{Shore-1979}. As a first step, the equations for the coefficient
$c_k(t)$ with $k\neq 0,1,N$ are solved under the approximation that the
slowly varying $c_0(t)$, $c_1(t)$ and $c_N(t)$ are constant, which yields
$c_k(t)\approx (1-\ee^{\ii\xi_k t})[V_{k1}c_1(t) + V_{kN}c_N(t)]/\xi_k$.
After inserting these approximate solutions into the equations for $c_0(t)$,
$c_1(t)$, $c_N(t)$ and neglecting all fast oscillating terms $\ee^{-\ii\xi_k
t}$ we obtain
\begin{equation}\label{effdyn}
    \ii\frac{\partial}{\partial t}\left(
      \begin{array}{c}
       c_1 \\
       c_0  \\
       c_N  \\
      \end{array}
    \right) \approx \left(
                \begin{array}{ccc}
                  \Delta_1 & V_{10} & \frac{1}{2}\Omega \\
                  V_{01} & 0 & V_{0N} \\
                  \frac{1}{2}\Omega & V_{N0} & \Delta_N \\
                \end{array}
              \right)\left(
                       \begin{array}{c}
                         c_1 \\
                         c_0 \\
                         c_N \\
                       \end{array}
                     \right)\,,
\end{equation}
where the detunings $\Delta_i$ and the frequency $\Omega$ are given by
\begin{equation}\nonumber
    \Delta_i = - \sum_k \frac{|V_{i0}|^2}{\xi_k}\qquad\Omega = - 2\sum_k\frac{V_{1k}V_{kN}}{\xi_k^2}\,.
\end{equation}
Because of the symmetry of the problem the detunings $\Delta_i$ vanish
identically~\cite{Shore-1979}. We discuss the dynamics of the state
$\ket{\varphi(t)}$ resulting from Eq.~(\ref{effdyn}) for the case $V_{01} =
V_{0N} = \nu/\sqrt{2}$ and $b_1 = b_{N-1} = b$.

If $N$ is odd then the system has a zero mode $\ket{\xi_0}$ and the dominant
contributions in the limit of weak couplings $b\rightarrow 0$ come from the
matrix elements $\nu\sim b$ since the second-order frequencies scale as
$\Omega\sim b^2$. The state evolves according to
\begin{equation}\nonumber
    \ket{\varphi(t)} = \cos^2\left(\frac{\nu t}{2}\right)\ket{1} - \frac{\ii\sin(\nu t)}{\sqrt{2}}\ket{\xi_0} - \sin^2\left(\frac{\nu
    t}{2}\right)\ket{N}
\end{equation}
and thus is transferred from site $1$ to $N$ after the time $\tau =
\pi/\nu\sim 1/b$. The (unnormalized) eigenstates of the matrix in
Eq.~(\ref{effdyn}) are given by $\ket{1} - \ket{N}$ and $\ket{1}
\pm\ket{\xi_0} + \ket{N}$ with eigenvalues $0$ and $\pm\nu$, respectively.
Thus the perturbation $V$ leads to strong mixing of the boundary states
$\ket{1}$, $\ket{N}$ with the zero mode $\ket{\xi_0}$ and lifts their
degeneracy with an energy splitting proportional to $b$.

\begin{figure}[t!]
\begin{center}
  \includegraphics[width = 210pt]{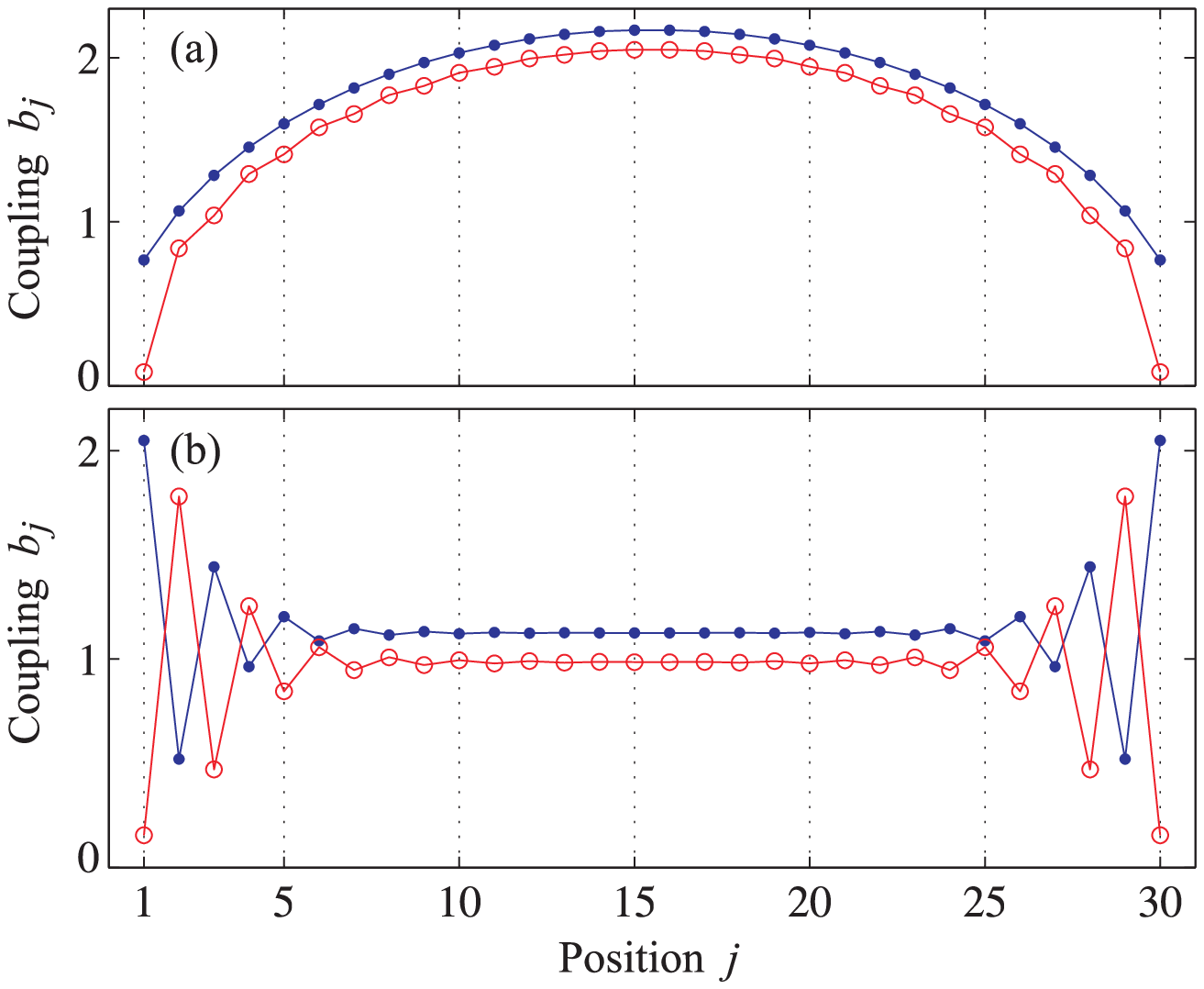}\vspace{10pt}\\
  \hspace{5pt}\includegraphics[width = 200pt]{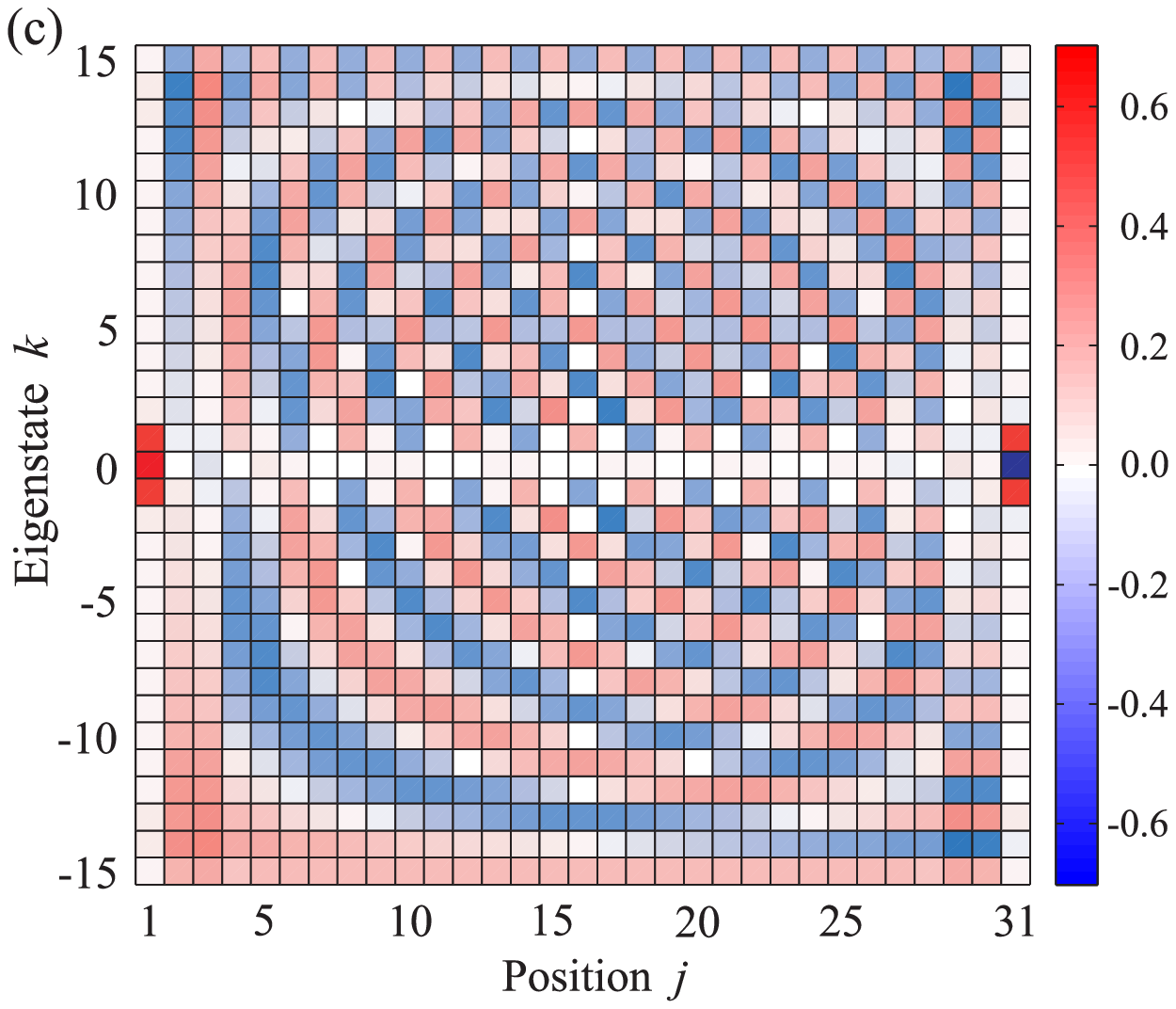}
  \caption{(Color online) Couplings $b_j$ between neighboring spins for perfect state
  transfer before (full dots $\bullet$) and after (empty dots $\circ$) adding boundary
  states to the spin chain. (a)~The outermost couplings $b_1$, $b_N$ of the spin chain
  with linear spectrum are reduced when boundary states are added. (b)~The couplings of the
  spin chain with inverted quadratic spectrum are approximately mirrored along the
  horizontal axis. (c)~Components of the eigenstates $\ket{\lambda_k}$ in the position
  basis $\{\ket{j}\}$ corresponding to the modified couplings in (b). The boundary
  states $\ket{\lambda_{-1}}$, $\ket{\lambda_{0}}$ and $\ket{\lambda_{1}}$ can be clearly seen.
  The parameters are set to $N=31$, $A=7$ and the $b_j$ are scaled by
  $1/25$ and $1/100$ in (a) and (b), respectively.}\label{couplings}
\end{center}
\end{figure}

If $N$ is even then the zero mode $\ket{\xi_0} $ is absent and
Eq.~(\ref{effdyn}) describes on-resonance Rabi oscillations between the
states $\ket{1}$ and $\ket{N}$ at the Rabi frequency $\Omega$. Accordingly
we have
\begin{equation}\nonumber
    \ket{\varphi(t)} = \cos\left(\frac{\Omega t}{2}\right)\ket{1} - \ii\sin\left(\frac{\Omega t}{2}\right)\ket{N}
\end{equation}
and state transfer takes place after the time $\tau = \pi/\Omega\sim 1/b^2$.
The eigenstates under the effect of the perturbation $V$ are $\ket{1} \pm
\ket{N}$ with eigenvalues $\pm\Omega$, and thus we have an energy splitting
proportional to $b^2$.

The perturbative approach is valid provided that
$\nu,\Omega\ll\xi_{\mathrm{min}}$, where $\xi_{\mathrm{min}}$ is the
non-zero eigenvalue with the smallest magnitude. In this regime the
contributions from high-frequency modes with $\xi_k\geq\xi_{\mathrm{min}}$
average out on the time scale of the state transfer. In particular, PST is
achieved only in the limit of vanishing couplings $b_1$ and $b_{N-1}$, which
is the main drawback of this approach. On the other hand, in this limit PST
is possible for arbitrary configurations of the inner couplings $b_j$ with
$j=2,\ldots,N-2$.


\section{Combined approach}

We combine the two approaches in order to endow modulated spin chains with
boundary states. State transfer then takes place mainly through the boundary
states, making it more robust to imperfections, and yet is perfect even for
finite couplings to the end spins. As seen previously, nearly zero
eigenvalues in the spectrum of the spin chain, i.e., eigenvalues
significantly smaller in magnitude than any other eigenvalue, are a
signature of boundary states. The crucial question is if the converse is
also true, i.e., if adding nearly zero eigenvalues is sufficient to
introduce boundary states. The answer is no---the presence or absence of
boundary states depends on the entire spectrum of the chain.

In order to show this we focus on spin chains with odd $N$; however, the
arguments are similar for even $N$. The condition for state transfer through
boundary states is $\nu\ll\xi_{\mathrm{min}}$, or equivalently
$b\ll\xi_{\mathrm{min}}$. To obtain a condition only on the spectrum we make
two observations: First, we find from the algorithm by de~Boor and Golub
that $b^2 = \sum_k w_k \lambda_k^2/\sum_k w_k$, which is the weighted
variance of the spectrum. Second, we notice that $\xi_{\mathrm{min}}$ is
identical to $\lambda_{\mathrm{min}}$ provided nearly zero eigenvalues are
excluded from the minimum. Therefore the condition for state transfer
through boundary states in terms of the eigenvalues $\lambda_k$ and the
weights $w_k = |\prod_{p\neq k}(\lambda_k-\lambda_p)|^{-1}$ reads
\begin{equation}\label{bspec}
    \frac{\sum_k w_k \lambda_k^2}{\sum_k w_k}\ll\lambda^2_{\mathrm{min}}\,.
\end{equation}

\begin{figure}[t!]
\begin{center}
  \includegraphics[width = 190pt]{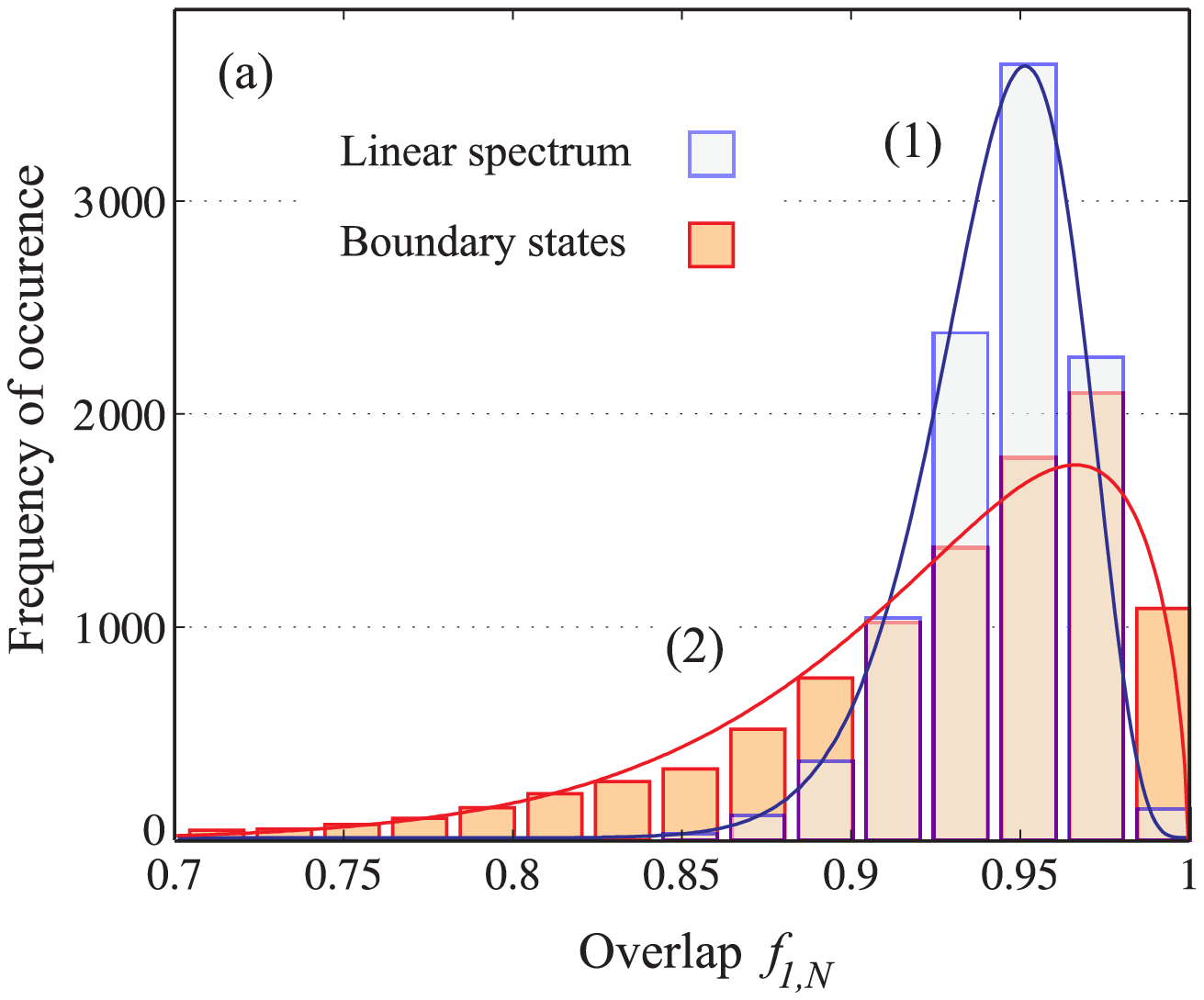}\vspace{10pt}\\
  \includegraphics[width = 190pt]{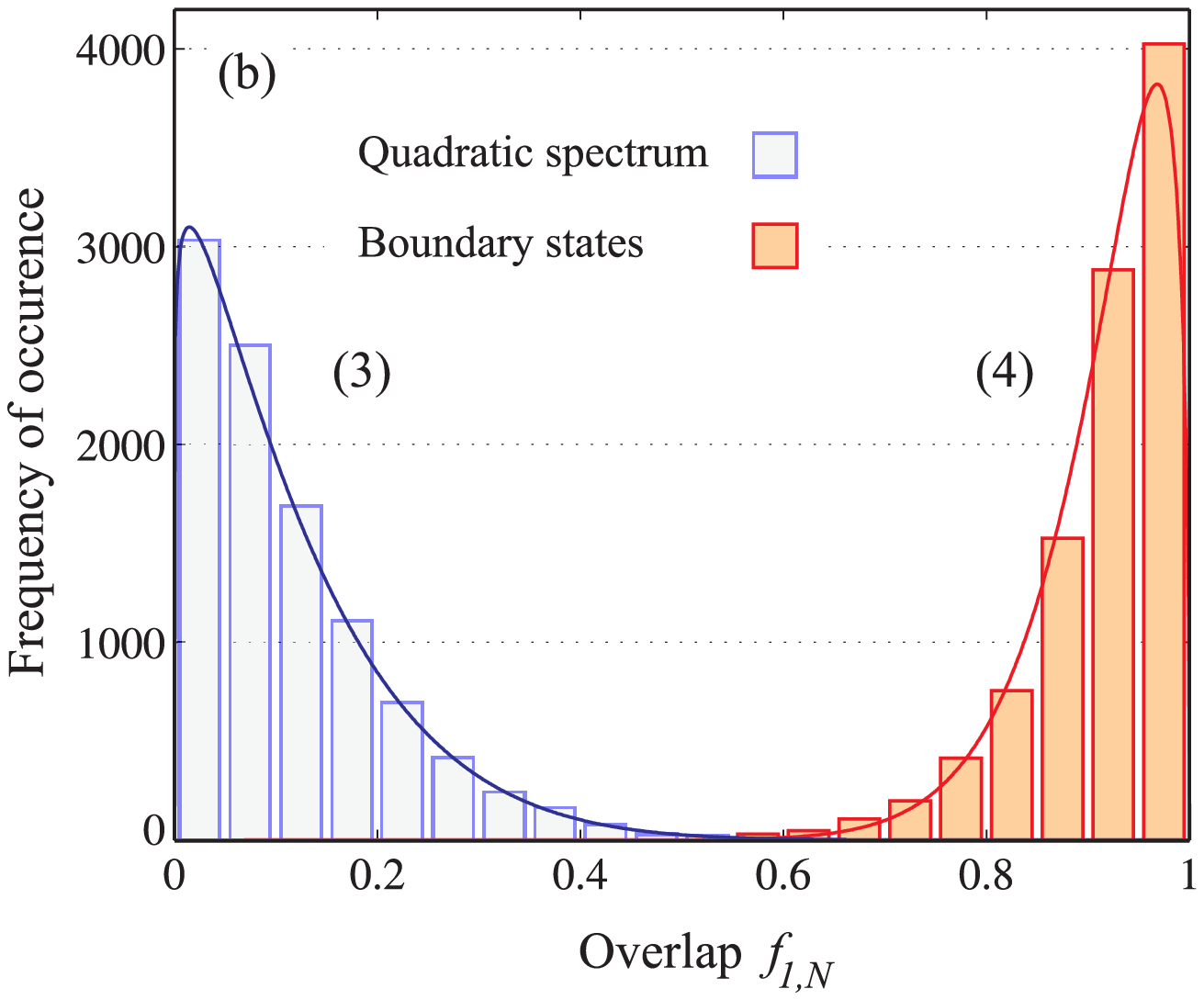}
  \caption{(Color online) The histogram and best-fit $\beta$ distribution of the overlap $f_{1,N}$
  for the same spin chains as in Fig.~1 with randomly perturbed couplings $b_j$. (a)~Adding
  boundary states to the chain with linear spectrum results in more instances of high-fidelity
  transport and a broader distribution of $f_{1,N}$. (b)~The average transport fidelity of the chain with
  inverted quadratic spectrum is significantly improved by adding boundary states.
  The best-fit $\beta$ distribution corresponding to the curves labeled (1)--(4) yields
  $\alpha = \{96.7,19.3,1.11,15.6\}$, $\beta = \{5.88,1.63,8.65,1.48\}$
  and the average $\mu =\{0.943,0.922,0.114,0.914\}$.}\label{hist}
\end{center}
\end{figure}

Even though Eq.~(\ref{bspec}) is the desired result we gain more insight by
discussing it on the basis of a stepwise linear spectrum. The specific
spectrum consist of three bands separated by gaps $\Gamma$, taking the role
of $\lambda_{\mathrm{min}}$. The $m$ nearly zero eigenvalues in the central
band and the $M$ eigenvalues in each of the peripheral bands have
inter-level spacing $\delta$ and $\Delta$, respectively. We estimate the
values of the weights $w_k$ to determine the regime of $\delta$, $\Gamma$,
$\Delta$ for which Eq.~(\ref{bspec}) is fulfilled. Noting that the weights
$w_k$ depend on the distance between the eigenvalues we find the scalings
\begin{equation}\nonumber\label{west}
    w_k^{\mathrm{cen}}\sim\frac{1}{\delta^m\Gamma^{2M}}\,,\qquad w_k^{\mathrm{per}}\sim\frac{1}{\Delta^{M}\Gamma^m(2\Gamma)^{M}}\,,
\end{equation}
where $w_k^{\mathrm{cen}}$ and $w_k^{\mathrm{per}}$ are the typical weights
for the eigenvalues $\lambda_k$ located in the central and the peripheral
bands, respectively. Now, Eq.~(\ref{bspec}) is fulfilled if contributions
from large eigenvalues in the peripheral bands to the weighted variance are
small, i.e., $w_k^{\mathrm{per}}\ll w_k^{\mathrm{cen}}$, or equivalently
\begin{equation}\nonumber\label{wcond}
    \left(\frac{\delta}{\Gamma}\right)^m\left(\frac{\Gamma}{\Delta}\right)^M\ll 1\,.
\end{equation}
Thus, eigenvalues in the central band lead to boundary states of the spin
chain in the regime $\delta\ll\Gamma$ and $\Delta\sim\Gamma$.

To illustrate the combined approach we consider two examples, namely, the
linear spectrum and an inverted quadratic spectrum. Nearly zero eigenvalues
are introduced by changing the original spectrum $\lambda_k$ to the shifted
spectrum $\tilde{\lambda}_k = \lambda_k - \mathrm{sgn}(\lambda_k)C$, where
$C<\lambda_{\mathrm{min}}$ is a constant. This shifting procedure is
applicable to all spectra and leaves the length of the spin chain unchanged.

The linear spectrum is given by $\lambda_k = A k$, where the index $k$ runs
from $-\frac{1}{2}(N-1)$ to $\frac{1}{2}(N-1)$, and $A\gg1$ is an arbitrary
odd integer so that $\tau = \pi/A$. After shifting with $C= A - 1$ the
spectrum contains two eigenvalues $\tilde{\lambda}_{\pm1} = \pm 1$ with
boundary states $\ket{\tilde{\lambda}_{\pm1}}$. Of the corresponding
couplings only $b_1$, $b_{N-1}$ are significantly reduced, as shown in
Fig.~\ref{couplings}(a). Both spin chains, with spectrum $\lambda_k$ and
$\tilde{\lambda}_k$, support PST and their transfer times are related by
$\tilde{\tau} = A\tau$.

The inverted quadratic spectrum is defined by $\lambda_k = k(N-1-|k|)$. The
corresponding couplings $b_j$ strongly oscillate toward the end of the chain
and are approximately constant in the center. The shifted spectrum with $C =
N - 3$ has two eigenvalues $\tilde{\lambda}_{\pm1} = \pm1$ with boundary
states $\ket{\tilde{\lambda}_{\pm1}}$. Both spin chains, with spectrum
$\lambda_k$ and $\tilde{\lambda}_k$, support PST with identical transfer
times. The couplings $b_j$ and the eigenstates $\ket{\lambda_k}$ are shown
in Figs.~\ref{couplings}(b) and~\ref{couplings}(c), respectively.

As an aside, spin chains with cosine spectrum $\lambda_k = 2\cos\left[\pi
k/(N+1)\right]$ and constant couplings $b_j = 1$ do not support PST but can
still be endowed with boundary states. The shifting procedure significantly
reduces the outermost couplings $b_1$, $b_{N-1}$ and leads to slight
oscillations of the inner couplings. This represents an alternative to the
modification of the cosine spectrum suggested in
Ref.~\cite{Karbach-PRA-2005} in order to achieve PST.

\subsection*{Random static imperfections}

We now turn to the performance of spin chains in the presence of static
imperfections in the couplings and show that boundary states improve the
transfer fidelity. For this purpose we numerically evaluate the transfer
fidelity of spin chains with randomized couplings $b^{\mathrm{rnd}}_j =
b_j(1 + R)$, where $R$ is a uniformly distributed random variable in the
interval $[-r,r]$. We use the overlap $f_{1,N}(\tau) = |\bra{N}\ee^{-\ii
H_N\tau}\ket{1}|$, taking values in the interval $[0,1]$, to assess the
performances of the chain. The overlap $f_{1,N}$ is related to the transfer
fidelity $F$ of the state transfer averaged over all input states on the
Bloch sphere by $F = \frac{1}{2} + \frac{1}{3}f_{1,N} +
\frac{1}{6}f_{1,N}^2$~\cite{Bose-CP-2007}. The transfer time $\tau$ is fixed
to the value of the perfectly engineered chain ($r = 0$), in which case
$f_{1,N}(\tau)=1$.

We reconsider the spin chains with linear and inverted quadratic spectrum
and compare the distribution of the overlap $f_{1,N}(\tau)$ with and without
boundary states, sampled over $10^4$ transfers. In addition, we fit a
$\beta$ distribution to the numerically obtained distribution of
$f_{1,N}(\tau)$. The $\beta$ distribution is defined on the interval $(0,1)$
by the probability density $P(x) =
[B(\alpha,\beta)]^{-1}\,x^{\alpha-1}(1-x)^{\beta-1}$, where $\alpha$ and
$\beta$ are two positive shape parameters and $B(\alpha,\beta)=\int_0^1\dd
t\,t^{\alpha-1}(1-t)^{\beta-1}$ is the $\beta$ function. The average value
is given by $\mu=\alpha/(\alpha+\beta)$ and the variance by
$\sigma^2=\alpha\beta/[(\alpha+\beta)^2(\alpha+\beta+1)]$.

Figure~\ref{hist} shows the distribution of the $f_{1,N}(\tau)$ for the
linear and inverted quadratic spectrum, as well as their shifted counterpart
with boundary states. In all cases the randomized couplings (at the level
$r=0.05$) result in reduced transfer fidelities. Both spin chains with
linear spectrum are remarkably resilient to imperfections; however, the
corresponding distributions of the overlap $f_{1,N}(\tau)$ differ
noticeably, as shown in Fig.~\ref{hist}(a). Boundary states lead to a
broader distribution of $f_{1,N}(\tau)$ and to substantially more instances
of high-fidelity state transfer. Spin chains with boundary states are
therefore advantageous in a scenario where varying couplings are caused by
imperfect fabrication (or tuning) and only high-quality chains, say with
$f_{1,N}(\tau)\geq0.98$, are selected. Note that the performance of the spin
chain with a linear spectrum can be further improved by increasing the
parameter $A$.

\begin{figure}[t!]
\begin{center}
  \hspace{-8pt}\includegraphics[width = 185pt]{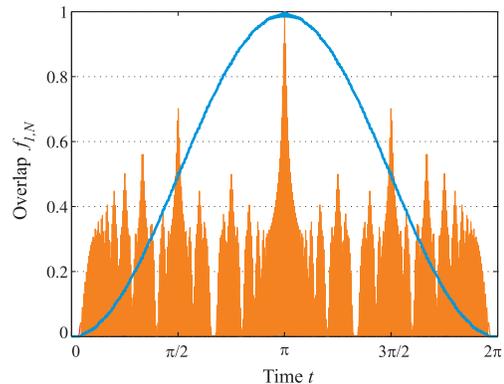}
  \caption{(Color online) The overlap $f_{1,N}$ as a function of time~$t$ for the
  spin chains with the inverted quadratic spectrum of Fig.~1(b). The irregular time dependence
  of $f_{1,N}$ (orange peaks) with high-frequency oscillations and a sharp peak at $t=\pi$ is
  smoothed out by adding boundary states, resulting in a regular sine dependence (blue curve)
  with small-scale oscillations (visible only as the thickness of the plotted curve).}\label{evol}
\end{center}
\end{figure}

The fidelity of chains with an inverted quadratic spectrum is clearly
improved by adding boundary states. The average fidelities and the
distributions of $f_{1,N}(\tau)$ differ significantly, as shown in
Fig.~\ref{hist}(b). This is readily explained by looking at the time
dependence of $f_{1,N}(t)$ in Fig.~\ref{evol}. Without boundary states the
overlap $f_{1,N}(t)$ carries out high-frequency oscillations and is sharply
peaked around the transfer time $\tau = \pi$. As a consequence, small
perturbations are likely to change the position of this peak and to reduce
the overlap $f_{1,N}(\tau)$ considerably. In contrast, $f_{1,N}(t)$ varies
smoothly if the time dependence is mainly determined by boundary states.

Our observations made above hold true for a wide range of parameters,
specifically, for different levels of imperfection $r$. This is to be
expected since all spin chains with boundary states are described by the
same effective two- or three-level system and therefore exhibit similar
behavior. Moreover, random imperfections of the spin chain, represented by
$\hat{H}_{\mathrm{rnd}}$, lead to lowest-order corrections of the states
$\ket{\lambda_k}$ and their energies $\lambda_k$ proportional to
$\sum_p\bra{\lambda_k}\hat{H}_{\mathrm{rnd}}\ket{\lambda_p}/(\lambda_k-\lambda_p)$
and $\bra{\lambda_k}\hat{H}_{\mathrm{rnd}}\ket{\lambda_k}$, respectively.
The relevant matrix elements are small for boundary states as long as the
perturbation $\hat{H}_{\mathrm{rnd}}$ mainly affects the inner part of the
chain, and mixing of boundary states with high-frequency states
$\ket{\lambda_p}$ is suppressed by $1/\lambda_p$. However, if the outermost
couplings are affected by the perturbation, strong mixing of boundary states
may occur.


\section{Conclusions}

In summary, we have presented a comprehensive approach for designing spin
chains suitable for high-fidelity state transfer. We have combined the two
strategies to achieve PST based on modulated couplings and weakly coupled
end spins with boundary states. This allows us to exploit their respective
advantages, namely, PST for finite couplings and resilience to
imperfections. We have shown that a large class of spin chains can be
endowed with boundary states by modifying their energy spectrum, provided it
fulfills the condition in Eq.~(\ref{bspec}).

We have evaluated the performance of different spin chains assuming that the
couplings between spins are affected by random static imperfections. We saw
that adding boundary states to spin chains significantly changes the
distribution of their transfer fidelities. Depending on the specific chain,
either the number of high-fidelity transfers or the average transfer
fidelity are increased. Part of this increase is explained by the smooth
dependence of the transfer fidelity on the evolution time if transfer is
achieved through boundary states.

The results for single-qubit transfer can be easily extended to registers of
qubits. Adding several nearly zero eigenvalues to the energy spectrum
results in boundary states localized over a few sites at both ends of the
spin chain. As there are many possible configurations for spin chains
supporting PST, e.g., adapted to a specific physical system, we provide a
graphical user interface for designing spin chains (\textsc{spinGUIn}) as
Supplemental Material~\cite{supplemental}.

\vspace{-8pt}


\acknowledgments

M.B., K.F. and W.B. acknowledge financial support from the German Research
Foundation (DFG) through SFB 767 and the Swiss National Science Foundation
(SNSF) through Project PBSKP2/130366.

\bibliography{spinGUIn2c}

\begin{thebibliography}{26}%
\makeatletter
\providecommand \@ifxundefined [1]{%
 \@ifx{#1\undefined}
}%
\providecommand \@ifnum [1]{%
 \ifnum #1\expandafter \@firstoftwo
 \else \expandafter \@secondoftwo
 \fi
}%
\providecommand \@ifx [1]{%
 \ifx #1\expandafter \@firstoftwo
 \else \expandafter \@secondoftwo
 \fi
}%
\providecommand \natexlab [1]{#1}%
\providecommand \enquote  [1]{``#1''}%
\providecommand \bibnamefont  [1]{#1}%
\providecommand \bibfnamefont [1]{#1}%
\providecommand \citenamefont [1]{#1}%
\providecommand \href@noop [0]{\@secondoftwo}%
\providecommand \href [0]{\begingroup \@sanitize@url \@href}%
\providecommand \@href[1]{\@@startlink{#1}\@@href}%
\providecommand \@@href[1]{\endgroup#1\@@endlink}%
\providecommand \@sanitize@url [0]{\catcode `\\12\catcode `\$12\catcode
  `\&12\catcode `\#12\catcode `\^12\catcode `\_12\catcode `\%12\relax}%
\providecommand \@@startlink[1]{}%
\providecommand \@@endlink[0]{}%
\providecommand \url  [0]{\begingroup\@sanitize@url \@url }%
\providecommand \@url [1]{\endgroup\@href {#1}{\urlprefix }}%
\providecommand \urlprefix  [0]{URL }%
\providecommand \Eprint [0]{\href }%
\providecommand \doibase [0]{http://dx.doi.org/}%
\providecommand \selectlanguage [0]{\@gobble}%
\providecommand \bibinfo  [0]{\@secondoftwo}%
\providecommand \bibfield  [0]{\@secondoftwo}%
\providecommand \translation [1]{[#1]}%
\providecommand \BibitemOpen [0]{}%
\providecommand \bibitemStop [0]{}%
\providecommand \bibitemNoStop [0]{.\EOS\space}%
\providecommand \EOS [0]{\spacefactor3000\relax}%
\providecommand \BibitemShut  [1]{\csname bibitem#1\endcsname}%
\let\auto@bib@innerbib\@empty
\bibitem [{\citenamefont {Bose}(2007)}]{Bose-CP-2007}%
  \BibitemOpen
  \bibfield  {author} {\bibinfo {author} {\bibfnamefont {S.}~\bibnamefont
  {Bose}},\ }\bibfield  {title} {\enquote {\bibinfo {title} {Quantum
  communication through spin chain dynamics: an introductory overview},}\
  }\href@noop {} {\bibfield  {journal} {\bibinfo  {journal} {Contemp. Phys.}\
  }\textbf {\bibinfo {volume} {48}},\ \bibinfo {pages} {13} (\bibinfo {year}
  {2007})}\BibitemShut {NoStop}%
\bibitem [{\citenamefont {Kay}(2010)}]{Kay-IJQI-2010}%
  \BibitemOpen
  \bibfield  {author} {\bibinfo {author} {\bibfnamefont {A.}~\bibnamefont
  {Kay}},\ }\bibfield  {title} {\enquote {\bibinfo {title} {Perfect, efficient,
  state transfer and its application as a constructive tool},}\ }\href@noop {}
  {\bibfield  {journal} {\bibinfo  {journal} {Int. J. Quantum Inf.}\ }\textbf
  {\bibinfo {volume} {8}},\ \bibinfo {pages} {641} (\bibinfo {year}
  {2010})}\BibitemShut {NoStop}%
\bibitem [{\citenamefont {Bose}(2003)}]{Bose-PRL-2003}%
  \BibitemOpen
  \bibfield  {author} {\bibinfo {author} {\bibfnamefont {S.}~\bibnamefont
  {Bose}},\ }\bibfield  {title} {\enquote {\bibinfo {title} {Quantum
  communication through an unmodulated spin chain},}\ }\href@noop {} {\bibfield
   {journal} {\bibinfo  {journal} {Phys. Rev. Lett.}\ }\textbf {\bibinfo
  {volume} {91}},\ \bibinfo {pages} {207901} (\bibinfo {year}
  {2003})}\BibitemShut {NoStop}%
\bibitem [{\citenamefont {Christandl}\ \emph {et~al.}(2004)\citenamefont
  {Christandl}, \citenamefont {Datta}, \citenamefont {Ekert},\ and\
  \citenamefont {Landahl}}]{Christandl-PRL-2004}%
  \BibitemOpen
  \bibfield  {author} {\bibinfo {author} {\bibfnamefont {M.}~\bibnamefont
  {Christandl}}, \bibinfo {author} {\bibfnamefont {N.}~\bibnamefont {Datta}},
  \bibinfo {author} {\bibfnamefont {A.}~\bibnamefont {Ekert}}, \ and\ \bibinfo
  {author} {\bibfnamefont {A.~J.}\ \bibnamefont {Landahl}},\ }\bibfield
  {title} {\enquote {\bibinfo {title} {Perfect state transfer in quantum spin
  networks},}\ }\href@noop {} {\bibfield  {journal} {\bibinfo  {journal} {Phys.
  Rev. Lett.}\ }\textbf {\bibinfo {volume} {92}},\ \bibinfo {pages} {187902}
  (\bibinfo {year} {2004})}\BibitemShut {NoStop}%
\bibitem [{\citenamefont {W\'ojcik}\ \emph {et~al.}(2005)\citenamefont
  {W\'ojcik}, \citenamefont {\L{}uczak}, \citenamefont {Kurzy\'{n}ski},
  \citenamefont {Grudka}, \citenamefont {Gdala},\ and\ \citenamefont
  {Bednarska}}]{Wojcik-PRA-2005}%
  \BibitemOpen
  \bibfield  {author} {\bibinfo {author} {\bibfnamefont {A.}~\bibnamefont
  {W\'ojcik}}, \bibinfo {author} {\bibfnamefont {T.}~\bibnamefont {\L{}uczak}},
  \bibinfo {author} {\bibfnamefont {P.}~\bibnamefont {Kurzy\'{n}ski}}, \bibinfo
  {author} {\bibfnamefont {A.}~\bibnamefont {Grudka}}, \bibinfo {author}
  {\bibfnamefont {T.}~\bibnamefont {Gdala}}, \ and\ \bibinfo {author}
  {\bibfnamefont {M.}~\bibnamefont {Bednarska}},\ }\bibfield  {title} {\enquote
  {\bibinfo {title} {Unmodulated spin chains as universal quantum wires},}\
  }\href@noop {} {\bibfield  {journal} {\bibinfo  {journal} {Phys. Rev. A}\
  }\textbf {\bibinfo {volume} {72}},\ \bibinfo {pages} {034303} (\bibinfo
  {year} {2005})}\BibitemShut {NoStop}%
\bibitem [{\citenamefont {Yung}\ and\ \citenamefont
  {Bose}(2005)}]{Yung-PRA-2005}%
  \BibitemOpen
  \bibfield  {author} {\bibinfo {author} {\bibfnamefont {M.-H.}\ \bibnamefont
  {Yung}}\ and\ \bibinfo {author} {\bibfnamefont {S.}~\bibnamefont {Bose}},\
  }\bibfield  {title} {\enquote {\bibinfo {title} {Perfect state transfer,
  effective gates, and entanglement generation in engineered bosonic and
  fermionic networks},}\ }\href@noop {} {\bibfield  {journal} {\bibinfo
  {journal} {Phys. Rev. A}\ }\textbf {\bibinfo {volume} {71}},\ \bibinfo
  {pages} {032310} (\bibinfo {year} {2005})}\BibitemShut {NoStop}%
\bibitem [{\citenamefont {Clark}\ \emph {et~al.}(2005)\citenamefont {Clark},
  \citenamefont {Alves},\ and\ \citenamefont {Jaksch}}]{Clark-NJP-2005}%
  \BibitemOpen
  \bibfield  {author} {\bibinfo {author} {\bibfnamefont {S.~R.}\ \bibnamefont
  {Clark}}, \bibinfo {author} {\bibfnamefont {C.~Moura}\ \bibnamefont {Alves}},
  \ and\ \bibinfo {author} {\bibfnamefont {D.}~\bibnamefont {Jaksch}},\
  }\bibfield  {title} {\enquote {\bibinfo {title} {Efficient generation of
  graph states for quantum computation},}\ }\href@noop {} {\bibfield  {journal}
  {\bibinfo  {journal} {New J. Phys.}\ }\textbf {\bibinfo {volume} {7}},\
  \bibinfo {pages} {124} (\bibinfo {year} {2005})}\BibitemShut {NoStop}%
\bibitem [{\citenamefont {Clark}\ \emph {et~al.}(2007)\citenamefont {Clark},
  \citenamefont {Klein}, \citenamefont {Bruderer},\ and\ \citenamefont
  {Jaksch}}]{Clark-NJP-2007}%
  \BibitemOpen
  \bibfield  {author} {\bibinfo {author} {\bibfnamefont {S.~R.}\ \bibnamefont
  {Clark}}, \bibinfo {author} {\bibfnamefont {A.}~\bibnamefont {Klein}},
  \bibinfo {author} {\bibfnamefont {M.}~\bibnamefont {Bruderer}}, \ and\
  \bibinfo {author} {\bibfnamefont {D.}~\bibnamefont {Jaksch}},\ }\bibfield
  {title} {\enquote {\bibinfo {title} {Graph state generation with noisy
  mirror-inverting spin chains},}\ }\href@noop {} {\bibfield  {journal}
  {\bibinfo  {journal} {New J. Phys.}\ }\textbf {\bibinfo {volume} {9}},\
  \bibinfo {pages} {202} (\bibinfo {year} {2007})}\BibitemShut {NoStop}%
\bibitem [{\citenamefont {Yao}\ \emph {et~al.}(2011)\citenamefont {Yao},
  \citenamefont {Jiang}, \citenamefont {Gorshkov}, \citenamefont {Gong},
  \citenamefont {Zhai}, \citenamefont {Duan},\ and\ \citenamefont
  {Lukin}}]{Yao-PRL-2011}%
  \BibitemOpen
  \bibfield  {author} {\bibinfo {author} {\bibfnamefont {N.~Y.}\ \bibnamefont
  {Yao}}, \bibinfo {author} {\bibfnamefont {L.}~\bibnamefont {Jiang}}, \bibinfo
  {author} {\bibfnamefont {A.~V.}\ \bibnamefont {Gorshkov}}, \bibinfo {author}
  {\bibfnamefont {Z.-X.}\ \bibnamefont {Gong}}, \bibinfo {author}
  {\bibfnamefont {A.}~\bibnamefont {Zhai}}, \bibinfo {author} {\bibfnamefont
  {L.-M.}\ \bibnamefont {Duan}}, \ and\ \bibinfo {author} {\bibfnamefont
  {M.~D.}\ \bibnamefont {Lukin}},\ }\bibfield  {title} {\enquote {\bibinfo
  {title} {Robust quantum state transfer in random unpolarized spin chains},}\
  }\href@noop {} {\bibfield  {journal} {\bibinfo  {journal} {Phys. Rev. Lett.}\
  }\textbf {\bibinfo {volume} {106}},\ \bibinfo {pages} {040505} (\bibinfo
  {year} {2011})}\BibitemShut {NoStop}%
\bibitem [{\citenamefont {Lyakhov}\ and\ \citenamefont
  {Bruder}(2005)}]{Lyakhov-NJP-2005}%
  \BibitemOpen
  \bibfield  {author} {\bibinfo {author} {\bibfnamefont {A.}~\bibnamefont
  {Lyakhov}}\ and\ \bibinfo {author} {\bibfnamefont {C.}~\bibnamefont
  {Bruder}},\ }\bibfield  {title} {\enquote {\bibinfo {title} {Quantum state
  transfer in arrays of flux qubits},}\ }\href@noop {} {\bibfield  {journal}
  {\bibinfo  {journal} {New J. Phys.}\ }\textbf {\bibinfo {volume} {7}},\
  \bibinfo {pages} {181} (\bibinfo {year} {2005})}\BibitemShut {NoStop}%
\bibitem [{\citenamefont {Lieb}\ \emph {et~al.}(1961)\citenamefont {Lieb},
  \citenamefont {Schultz},\ and\ \citenamefont {Mattis}}]{Lieb-AP-1961}%
  \BibitemOpen
  \bibfield  {author} {\bibinfo {author} {\bibfnamefont {E.}~\bibnamefont
  {Lieb}}, \bibinfo {author} {\bibfnamefont {T.}~\bibnamefont {Schultz}}, \
  and\ \bibinfo {author} {\bibfnamefont {D.}~\bibnamefont {Mattis}},\
  }\bibfield  {title} {\enquote {\bibinfo {title} {Two soluble models of an
  antiferromagnetic chain},}\ }\href@noop {} {\bibfield  {journal} {\bibinfo
  {journal} {Ann. Phys.}\ }\textbf {\bibinfo {volume} {16}},\ \bibinfo {pages}
  {407} (\bibinfo {year} {1961})}\BibitemShut {NoStop}%
\bibitem [{\citenamefont {Eberly}\ \emph {et~al.}(1977)\citenamefont {Eberly},
  \citenamefont {Shore}, \citenamefont {Bia\l{}ynicka-Birula},\ and\
  \citenamefont {Bia\l{}ynicki-Birula}}]{Eberly-PRA-1977}%
  \BibitemOpen
  \bibfield  {author} {\bibinfo {author} {\bibfnamefont {J.~H.}\ \bibnamefont
  {Eberly}}, \bibinfo {author} {\bibfnamefont {B.~W.}\ \bibnamefont {Shore}},
  \bibinfo {author} {\bibfnamefont {Z.}~\bibnamefont {Bia\l{}ynicka-Birula}}, \
  and\ \bibinfo {author} {\bibfnamefont {I.}~\bibnamefont
  {Bia\l{}ynicki-Birula}},\ }\bibfield  {title} {\enquote {\bibinfo {title}
  {Coherent dynamics of {$N$}-level atoms and molecules. {I}. {N}umerical
  experiments},}\ }\href@noop {} {\bibfield  {journal} {\bibinfo  {journal}
  {Phys. Rev. A}\ }\textbf {\bibinfo {volume} {16}},\ \bibinfo {pages} {2038}
  (\bibinfo {year} {1977})}\BibitemShut {NoStop}%
\bibitem [{\citenamefont {Bia\l{}ynicka-Birula}\ \emph
  {et~al.}(1977)\citenamefont {Bia\l{}ynicka-Birula}, \citenamefont
  {Bia\l{}ynicki-Birula}, \citenamefont {Eberly},\ and\ \citenamefont
  {Shore}}]{Bialynicka-PRA-1977}%
  \BibitemOpen
  \bibfield  {author} {\bibinfo {author} {\bibfnamefont {Z.}~\bibnamefont
  {Bia\l{}ynicka-Birula}}, \bibinfo {author} {\bibfnamefont {I.}~\bibnamefont
  {Bia\l{}ynicki-Birula}}, \bibinfo {author} {\bibfnamefont {J.~H.}\
  \bibnamefont {Eberly}}, \ and\ \bibinfo {author} {\bibfnamefont {B.~W.}\
  \bibnamefont {Shore}},\ }\bibfield  {title} {\enquote {\bibinfo {title}
  {Coherent dynamics of {$N$}-level atoms and molecules. {II}. {A}nalytic
  solutions},}\ }\href@noop {} {\bibfield  {journal} {\bibinfo  {journal}
  {Phys. Rev. A}\ }\textbf {\bibinfo {volume} {16}},\ \bibinfo {pages} {2048}
  (\bibinfo {year} {1977})}\BibitemShut {NoStop}%
\bibitem [{\citenamefont {Cook}\ and\ \citenamefont
  {Shore}(1979)}]{Cook-PRA-1979}%
  \BibitemOpen
  \bibfield  {author} {\bibinfo {author} {\bibfnamefont {Richard~J.}\
  \bibnamefont {Cook}}\ and\ \bibinfo {author} {\bibfnamefont {Bruce~W.}\
  \bibnamefont {Shore}},\ }\bibfield  {title} {\enquote {\bibinfo {title}
  {Coherent dynamics of {$N$}-level atoms and molecules. {III}. {A}n
  analytically soluble periodic case},}\ }\href@noop {} {\bibfield  {journal}
  {\bibinfo  {journal} {Phys. Rev. A}\ }\textbf {\bibinfo {volume} {20}},\
  \bibinfo {pages} {539} (\bibinfo {year} {1979})}\BibitemShut {NoStop}%
\bibitem [{\citenamefont {Shore}\ and\ \citenamefont
  {Cook}(1979)}]{Shore-1979}%
  \BibitemOpen
  \bibfield  {author} {\bibinfo {author} {\bibfnamefont {Bruce~W.}\
  \bibnamefont {Shore}}\ and\ \bibinfo {author} {\bibfnamefont {Richard~J.}\
  \bibnamefont {Cook}},\ }\bibfield  {title} {\enquote {\bibinfo {title}
  {Coherent dynamics of {$N$}-level atoms and molecules. {IV}. {T}wo- and
  three-level behavior},}\ }\href@noop {} {\bibfield  {journal} {\bibinfo
  {journal} {Phys. Rev. A}\ }\textbf {\bibinfo {volume} {20}},\ \bibinfo
  {pages} {1958} (\bibinfo {year} {1979})}\BibitemShut {NoStop}%
\bibitem [{\citenamefont {W\'ojcik}\ \emph {et~al.}(2007)\citenamefont
  {W\'ojcik}, \citenamefont {\L{}uczak}, \citenamefont
  {Kurzy\ifmmode~\acute{n}\else \'{n}\fi{}ski}, \citenamefont {Grudka},
  \citenamefont {Gdala},\ and\ \citenamefont {Bednarska}}]{Wojcik-PRA-2007}%
  \BibitemOpen
  \bibfield  {author} {\bibinfo {author} {\bibfnamefont {A.}~\bibnamefont
  {W\'ojcik}}, \bibinfo {author} {\bibfnamefont {T.}~\bibnamefont {\L{}uczak}},
  \bibinfo {author} {\bibfnamefont {P.}~\bibnamefont
  {Kurzy\ifmmode~\acute{n}\else \'{n}\fi{}ski}}, \bibinfo {author}
  {\bibfnamefont {A.}~\bibnamefont {Grudka}}, \bibinfo {author} {\bibfnamefont
  {T.}~\bibnamefont {Gdala}}, \ and\ \bibinfo {author} {\bibfnamefont
  {M.}~\bibnamefont {Bednarska}},\ }\bibfield  {title} {\enquote {\bibinfo
  {title} {Multiuser quantum communication networks},}\ }\href@noop {}
  {\bibfield  {journal} {\bibinfo  {journal} {Phys. Rev. A}\ }\textbf {\bibinfo
  {volume} {75}},\ \bibinfo {pages} {022330} (\bibinfo {year}
  {2007})}\BibitemShut {NoStop}%
\bibitem [{\citenamefont {{Zwick}}\ \emph {et~al.}(2012)\citenamefont
  {{Zwick}}, \citenamefont {{Alvarez}}, \citenamefont {{Stolze}},\ and\
  \citenamefont {{Osenda}}}]{Zwick-PRA-2012}%
  \BibitemOpen
  \bibfield  {author} {\bibinfo {author} {\bibfnamefont {A.}~\bibnamefont
  {{Zwick}}}, \bibinfo {author} {\bibfnamefont {G.~A.}\ \bibnamefont
  {{Alvarez}}}, \bibinfo {author} {\bibfnamefont {J.}~\bibnamefont {{Stolze}}},
  \ and\ \bibinfo {author} {\bibfnamefont {O.}~\bibnamefont {{Osenda}}},\
  }\bibfield  {title} {\enquote {\bibinfo {title} {Spin chains for robust state
  transfer: Modified boundary couplings versus completely engineered chains},}\
  }\href@noop {} {\bibfield  {journal} {\bibinfo  {journal} {Phys. Rev. A}\
  }\textbf {\bibinfo {volume} {85}},\ \bibinfo {pages} {012318} (\bibinfo
  {year} {2012})}\BibitemShut {NoStop}%
\bibitem [{\citenamefont {de~Boor}\ and\ \citenamefont
  {Golub}(1978)}]{deBoor-LAA-1978}%
  \BibitemOpen
  \bibfield  {author} {\bibinfo {author} {\bibfnamefont {C.}~\bibnamefont
  {de~Boor}}\ and\ \bibinfo {author} {\bibfnamefont {G.~H.}\ \bibnamefont
  {Golub}},\ }\bibfield  {title} {\enquote {\bibinfo {title} {The numerically
  stable reconstruction of a {J}acobi matrix from spectral data},}\ }\href@noop
  {} {\bibfield  {journal} {\bibinfo  {journal} {Linear Algebr. Appl.}\
  }\textbf {\bibinfo {volume} {21}},\ \bibinfo {pages} {245} (\bibinfo {year}
  {1978})}\BibitemShut {NoStop}%
\bibitem [{sup()}]{supplemental}%
  \BibitemOpen
  \href@noop {} {}\bibinfo {note} {The source file of this article contains the
  \textsc{matlab} program files of \textsc{spinGUIn} including the implemented
  algorithm by de~Boor and Golub}\BibitemShut {NoStop}%
\bibitem [{\citenamefont {De~Chiara}\ \emph {et~al.}(2005)\citenamefont
  {De~Chiara}, \citenamefont {Rossini}, \citenamefont {Montangero},\ and\
  \citenamefont {Fazio}}]{DeChiara-PRA-2005}%
  \BibitemOpen
  \bibfield  {author} {\bibinfo {author} {\bibfnamefont {G.}~\bibnamefont
  {De~Chiara}}, \bibinfo {author} {\bibfnamefont {D.}~\bibnamefont {Rossini}},
  \bibinfo {author} {\bibfnamefont {S.}~\bibnamefont {Montangero}}, \ and\
  \bibinfo {author} {\bibfnamefont {R.}~\bibnamefont {Fazio}},\ }\bibfield
  {title} {\enquote {\bibinfo {title} {From perfect to fractal transmission in
  spin chains},}\ }\href@noop {} {\bibfield  {journal} {\bibinfo  {journal}
  {Phys. Rev. A}\ }\textbf {\bibinfo {volume} {72}},\ \bibinfo {pages} {012323}
  (\bibinfo {year} {2005})}\BibitemShut {NoStop}%
\bibitem [{\citenamefont {Zwick}\ \emph {et~al.}(2011)\citenamefont {Zwick},
  \citenamefont {\'Alvarez}, \citenamefont {Stolze},\ and\ \citenamefont
  {Osenda}}]{Zwick-PRA-2011}%
  \BibitemOpen
  \bibfield  {author} {\bibinfo {author} {\bibfnamefont {A.}~\bibnamefont
  {Zwick}}, \bibinfo {author} {\bibfnamefont {G.~A.}\ \bibnamefont
  {\'Alvarez}}, \bibinfo {author} {\bibfnamefont {J.}~\bibnamefont {Stolze}}, \
  and\ \bibinfo {author} {\bibfnamefont {O.}~\bibnamefont {Osenda}},\
  }\bibfield  {title} {\enquote {\bibinfo {title} {Robustness of spin-coupling
  distributions for perfect quantum state transfer},}\ }\href@noop {}
  {\bibfield  {journal} {\bibinfo  {journal} {Phys. Rev. A}\ }\textbf {\bibinfo
  {volume} {84}},\ \bibinfo {pages} {022311} (\bibinfo {year}
  {2011})}\BibitemShut {NoStop}%
\bibitem [{\citenamefont {Wang}\ \emph {et~al.}(2011)\citenamefont {Wang},
  \citenamefont {Shuang},\ and\ \citenamefont {Rabitz}}]{Wang-PRA-2011}%
  \BibitemOpen
  \bibfield  {author} {\bibinfo {author} {\bibfnamefont {Y.}~\bibnamefont
  {Wang}}, \bibinfo {author} {\bibfnamefont {F.}~\bibnamefont {Shuang}}, \ and\
  \bibinfo {author} {\bibfnamefont {H.}~\bibnamefont {Rabitz}},\ }\bibfield
  {title} {\enquote {\bibinfo {title} {All possible coupling schemes in
  $\mathit{XY}$ spin chains for perfect state transfer},}\ }\href@noop {}
  {\bibfield  {journal} {\bibinfo  {journal} {Phys. Rev. A}\ }\textbf {\bibinfo
  {volume} {84}},\ \bibinfo {pages} {012307} (\bibinfo {year}
  {2011})}\BibitemShut {NoStop}%
\bibitem [{\citenamefont {Sussman-Fort}(1982)}]{Sussman-JFI-1982}%
  \BibitemOpen
  \bibfield  {author} {\bibinfo {author} {\bibfnamefont {S.~E.}\ \bibnamefont
  {Sussman-Fort}},\ }\bibfield  {title} {\enquote {\bibinfo {title} {The
  reconstruction of bordered-diagonal and {J}acobi matrices from spectral
  data},}\ }\href@noop {} {\bibfield  {journal} {\bibinfo  {journal} {J.
  Franklin Inst.}\ }\textbf {\bibinfo {volume} {314}},\ \bibinfo {pages} {271}
  (\bibinfo {year} {1982})}\BibitemShut {NoStop}%
\bibitem [{\citenamefont {Hochstadt}(1974)}]{Hochstadt-LAA-1974}%
  \BibitemOpen
  \bibfield  {author} {\bibinfo {author} {\bibfnamefont {H.}~\bibnamefont
  {Hochstadt}},\ }\bibfield  {title} {\enquote {\bibinfo {title} {On the
  construction of a {J}acobi matrix from spectral data},}\ }\href@noop {}
  {\bibfield  {journal} {\bibinfo  {journal} {Linear Algebr. Appl.}\ }\textbf
  {\bibinfo {volume} {8}},\ \bibinfo {pages} {435} (\bibinfo {year}
  {1974})}\BibitemShut {NoStop}%
\bibitem [{\citenamefont {Langhoff}\ \emph {et~al.}(1972)\citenamefont
  {Langhoff}, \citenamefont {Epstein},\ and\ \citenamefont
  {Karplus}}]{Langhoff-RMP-1972}%
  \BibitemOpen
  \bibfield  {author} {\bibinfo {author} {\bibfnamefont {P.~W.}\ \bibnamefont
  {Langhoff}}, \bibinfo {author} {\bibfnamefont {S.~T.}\ \bibnamefont
  {Epstein}}, \ and\ \bibinfo {author} {\bibfnamefont {M.}~\bibnamefont
  {Karplus}},\ }\bibfield  {title} {\enquote {\bibinfo {title} {Aspects of
  time-dependent perturbation theory},}\ }\href@noop {} {\bibfield  {journal}
  {\bibinfo  {journal} {Rev. Mod. Phys.}\ }\textbf {\bibinfo {volume} {44}},\
  \bibinfo {pages} {602} (\bibinfo {year} {1972})}\BibitemShut {NoStop}%
\bibitem [{\citenamefont {Karbach}\ and\ \citenamefont
  {Stolze}(2005)}]{Karbach-PRA-2005}%
  \BibitemOpen
  \bibfield  {author} {\bibinfo {author} {\bibfnamefont {P.}~\bibnamefont
  {Karbach}}\ and\ \bibinfo {author} {\bibfnamefont {J.}~\bibnamefont
  {Stolze}},\ }\bibfield  {title} {\enquote {\bibinfo {title} {Spin chains as
  perfect quantum state mirrors},}\ }\href@noop {} {\bibfield  {journal}
  {\bibinfo  {journal} {Phys. Rev. A}\ }\textbf {\bibinfo {volume} {72}},\
  \bibinfo {pages} {030301} (\bibinfo {year} {2005})}\BibitemShut {NoStop}%
\end{thebibliography}%

\onecolumngrid

\newpage

\twocolumngrid

\section*{Supplemental material}\vspace{-5pt}

We here provide a graphical user interface for \textsc{matlab}, called
\textsc{spinGUIn}, that allows the user to generate and modify spin chains
for perfect state transfer. The files required to run \textsc{spinGUIn} are
\texttt{iepsolve.m}, \texttt{spinGUIn.fig} and \texttt{spinGUIn.p}, which
are added as supplemental material to this article or, alternatively, can be
downloaded from the web page
\texttt{http://cms.uni-konstanz.de/physik/belzig/ research/publications}. In
order to start \textsc{spinGUIn} one has to copy the files into the Matlab
current folder and run the function file \texttt{spinGUIn.p}. Furthermore,
the implemented algorithm by de~Boor and Golub (\texttt{iepsolve.m}) is
given below for quick copy and paste.

\begin{sidewaysfigure}
\hspace{-1cm} \vspace{-0.7cm}
\begin{center}
  \hspace{-2cm}\includegraphics[width = 650pt]{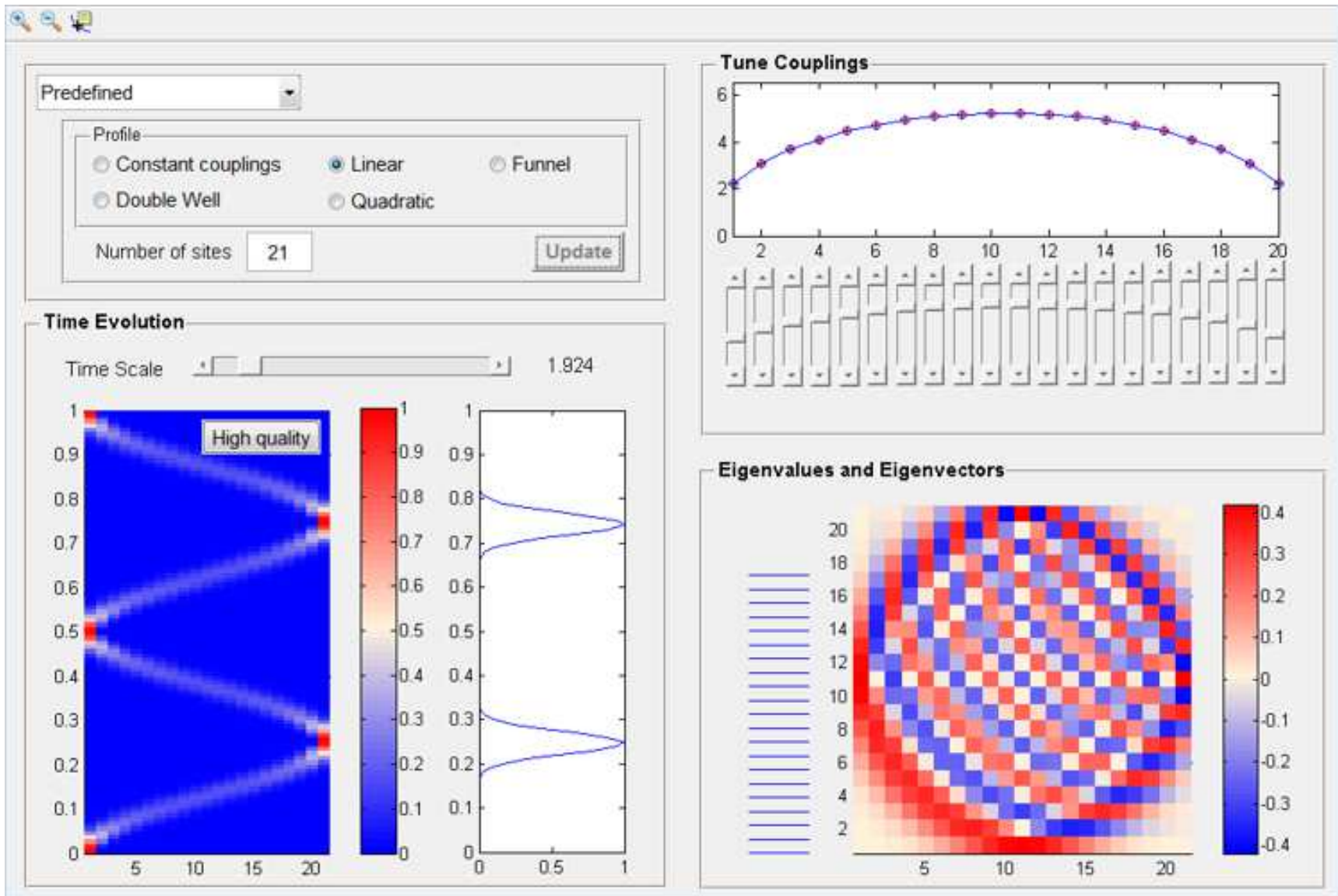}\\
  \hspace{-11cm}\parbox{13cm}{
  \caption{Screenshot of \textsc{spinGUIn}: The upper left panel is used to define
  the spin chain either via eigenvalues or spin couplings. The remaining
  panels show (i) the time-evolution of the state including the transfer
  fidelity (ii) the tunable couplings and (iii) the eigenvalues and eigenstates
  of the spin chain.}\label{fid}}
\end{center}
\end{sidewaysfigure}

As can be seen in Fig.~\ref{fid}, \textsc{spinGUIn} consist of four main
panels: The upper left panel is used to define the spin chain either via
eigenvalues or spin couplings. The remaining panels show (i) the time
evolution of the state including the transfer fidelity (ii) the tunable
couplings and (iii) the eigenvalues and eigenstates of the spin chain.

\onecolumngrid

\vspace{10pt}

\begin{verbatim}
function [a,b,mat] = iepsolve(spec)
%
% IEPSOLVE solves an inverse eigenvalue problem: The function constructs
% the persymmetric tridiagonal matrix of size N*N for a given set of N real
% non-degenerate eigenvalues.
%
% [A,B,MAT] = IEPSOLVE(SPEC)
%
% SPEC is the vector containing the eigenvalues (no ordering required).
%
% MAT is the complete persymmetric tridiagonal matrix.
% A is the vector containing the diagonal elements of MAT.
% B is the vector containing subdiagonal (superdiagonal) elements of MAT.
%
% Supplemental material to the article by M. Bruderer et al. (2011)
% 'Exploiting boundary states of imperfect spin chains for high-fidelity
%  state transfer'
%

% Determine the spectrum size N and the iteration maximum M
N = size(spec,2);
if rem(N,2)
    M = ceil(N/2);
else
    M = N/2;
end

% Shift and rescale the spectrum into the interval [-1,1]
mpos = (max(spec) + min(spec))/2;
spec = spec - mpos;
scale = max(abs(spec));
spec = spec/scale;

% Initialize the necessary arrays a_j, (b_j)^2, w_k and p_{j-1}
a = zeros(1,M);
bb = zeros(1,M);
w = zeros(1,N);
p1 = ones(1,N);

% Calculate the weights w_k from the spectrum
tmp = spec(2:N);
for  k = 1:N
    w(k) = abs(prod(1./(spec(k) - tmp)));
    tmp(k) = spec(k);
end
\end{verbatim}

\newpage

\pagestyle{empty}

\begin{verbatim}
% Calculate the components a_1, (b_1)^2 and the polynomial p_1
nom_a = sum(spec .* w);
den = sum(w);
a(1) = nom_a / den;
p = spec - a(1);
nom_b = sum(w .* p.^2);
bb(1) = nom_b / den;

% Calculate the remaining components a_j, (b_j)^2 and the polynomials p_j
for j = 2:M
    p2 = p1;
    p1 = p;
    nom_a = sum(spec .* w .* p1.^2);
    den = sum(w .* p1.^2);
    a(j) = nom_a / den;
    p = (spec - a(j)) .* p1 - bb(j-1) .* p2;
    nom_b = sum(w .* p.^2);
    bb(j) = nom_b / den;
end

% Compile and rescale the results using symmetry properties of the matrix
a = scale * a + mpos;
b = scale * sqrt(bb);
if rem(N,2)
    a = [a(1:M-1),fliplr(a)];
    b = [b(1:M-1),fliplr(b(1:M-1))];
else
    a = [a(1:M),fliplr(a(1:M))];
    b = [b(1:M-1),fliplr(b)];
end
mat =  diag(a) + diag(b,1) + diag(b,-1);
\end{verbatim}

\begin{figure}[h!]\vspace{3cm}
\begin{center}
  \includegraphics[width = 100pt]{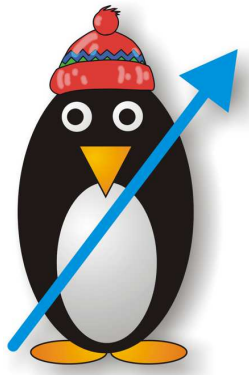}
  \caption{The authors' rendition of a \textsc{spinGUIn}}\label{spinguin}
\end{center}
\end{figure}

\end{document}